# ANONYMOUS AUCTION PROTOCOL BASED ON TIME-RELEASED ENCRYPTION ATOP CONSORTIUM BLOCKCHAIN

Jie Xiong and Qi Wang

Department of Computer Science, Jinan University, Guangzhou, China

*ABSTRACT*

*The Ethereum block chain as a decentralized platform is so successful that many applications deployed on it. However, for the inherent transparency properties and the lack of privacy, deploying a financial application on top of it is always a challenge. In this paper, we tackle this challenge and propose an anonymous sealed-bid auction protocol based on time-released encryption atop Consortium Block chain. We adopt a strict digital certificate-based identity mechanism of the consortium block chain to permit legitimate participants, and utilize the blind signature based on elliptic curve technology to allowing anonymous participation. Moreover, a timed release public key encryption algorithm is adopted to encrypt bids and prevent auctioneer from colluding with bidders. This is completely different from the method (zero-knowledge proof) used in other papers to prevent collusion between auctioneer and bidder. We provide a specific analysis of our protocol, which shows that our protocol meets anonymity and applicability.*

## 1. INTRODUCTION

Electronic auction is one of the basic businesses in electronic commerce [28], which is to transfer the real offline auction scenarios to the Internet. Thus they have the same basic components, that is, auction participants, auction rules and an arbitration institution. Among them, the auction participants include bidders and sellers (auctioneers). Auction rules refer to the principles which recognized and established by the auctioneer and bidder in the process of an auction. The arbitration institution is responsible for resolving disputes and conflicts during the auction. Online auctions have the advantages of low cost, wide range and high speed, which is more convenient and time-saving for participants
.
Traditionally, there are two types of auctions [1]:

1. Sealed-bid auction. This auction system requires that each bidder submits a bid price in sealed envelope and hands it to the auctioneer before the specified time. After the specified time, these bids can be opened by auctioneer and the winning bidder can be selected according to certain rules.

2. Open-bid auction. In this auction system, all bid values are disclosed, and bidders are allowed to submit bid more than once.

In this paper, our auction protocol is designed for sealed-bid auction of single-round bidding. Specifically, an auctioneer needs to purchase a batch of goods, a group of suppliers can provided goods and they submit bids. The auctioneer wants to buy these goods at the lowest price, each supplier hopes to get the trading opportunity to trade with the auctioneer. To facilitate this mechanism, the state-of-the-art solution requires a trusted third party (TTP) to host auction tasks to achieve the privacy of the participants and the fairness exchange. But TTP stores a lot of important information about users, so it is comes with potential threats from single-point attacks to collusion attacks all the time, also it is difficult to find a fully trusted institution to play such a role in reality.

Recently, many auction protocols were deployed on top of block chain. They take advantage of the decentralization and transparency properties of the block chain to get rid of the shortcomings which brought by the third party, for that everyone can check and verify the information on the ledger. In other words, there exists conflicts in preserving the privacy of the bids and trusting the auctioneer to compute the lowest bid privately on open block chain. In order to solve such challenges, cryptographic protocols can be utilized, such as secure multiparty computing (SMC), secret sharing, etc. But previous research has shown these protocols will make the scheme especially complicated, which leads to huge communication and computing overheads. There are also some literature [2], [3], that proposed to use zero-knowledge proof (ZKP) technique to prove that the auctioneer is legal in counting all the bids value and publishing the results of the auction, that is, preventing the auctioneer and the bidder from colluding. Especially for the ZKP, it takes a long time to generate proofs, and to deployment ZKP on smart contracts is particularly complicated. The data on the chain is publicly visible, so another challenge is that we cannot guarantee that the auctioneer won't decrypt the bid on the block chain during bidding time, and then secretly leaks the price, or disguises himself as a new bidder to participate in it.

According to the level of permission to join the chain, Block chains can be regarded as three types: Public Block chain, Private Block chain and Consortium Block chain. One of the most special features of the Consortium Block chain is that any entity node who wants to join the chain needs permission from the alliance. The Consortium Block chain can be regarded as a specific range of distributed TTP with high security and credibility. Therefore, it is suitable to announce rigorous auction activities with identity-based permitting mechanisms, such as limiting the attributes of participants.

In this paper, we present an anonymous auction protocol based on time-released encryption on Consortium Block chain. We utilize cryptographic primitives included time release encryption to guarantee the fairness and security, and blind signature to guarantee the bidder privacy. Specifically, we implement the following features:

1. Financial fairness. The auctioneer can only decrypt the bid after a certain time, so he cannot leak any information about bids to other participants who have not bid yet during the bidding period. And if the auctioneer aborts in the middle, as a punishment, his deposit will be distributed to other bidders. Also bidders will be disqualified if they quit halfway.

2. Non-repeatable bid. In an auction, the user can only bid once. If a bidder tries to bid multiple times in one task, the contract will check and cast off the message.

3.Bid privacy. The bid value will be encrypted. Bidders cannot know the bids submitted by the others before committing to their own. The auctioneer can only know all bids after a specified time.

4. Identity anonymity. Only the initiator of the auction (auctioneer) and CA know who participated in this auction, but no one can bind a certain bid to a unique identity. And the bidding behavior of the same user in multiple auctions cannot be linked to each other.

5. Public verifiable correctness. The data generated during the auction will be written on the chain, so the participating nodes can verify its correctness. Moreover, the final auction results will also be published on the block chain can be verified by anyone.

## 2. RELATED WORK

The electronic auction will find a center as an arbitration institution. This arbitration institution is generally a trusted third party (TTP) [4], [5], [6]. Usually, the instance of TTP can be an electronic bank, a certificate issuing authority, or a key distribution authority. Firstly, TTP publishes the auction rules, the deposit value of the auctioneer, bidding time periods and the time to open the result. If the legitimate users are interested, they can submit their bids to TTP in a certain format. In a sealed auction, this bid value should be hidden, after a period of time, TTP will give out the result or open the winning bidder. TTP is also responsible for resolving exceptions during the auction, such as someone quit halfway. Many centralized online auction researches [7], [8], [9] rely on a TTP, and they assume that TTP is semi-honest that he will not collude with the bidder. It is will-known that the third party stores too much sensitive data, masters too much power, it is impossible to trust him completely. In reality, numerous real word incidents reveal that the party might misbehave for self-interests [10], [11] privately, or some of attackers [12] can compromise its functionality.

In order to avoid the deficiencies brought about by centralization, many researches gradually turn to discussing the use of multiple centers to weaken the power of one center. For example, [13], [14], [15] propose multiple auction platforms (APs), they assume most of APs are honest. They get the auction results, which are calculated by multiple APs through SMC, secret sharing, etc. In [16], Brandt et al. propose using the announcement of encrypt binary bidding lists on a blackboard. It uses top-down, bottom-up and binary search techniques to interactively find the winner bid without revealing unnecessary information. In [17], [18], Abe et al. use homomorphic encryption, the mix and match technique; it proposes that the auction results can be jointly calculated in cipher text by each bidder in an interactive manner. Among them, message exchange is realized through secure channel, which abandons the center and guarantees the privacy of bids. However, multiple interactions between bidding nodes are required, it costs a lot of communication between nodes and greatly increases the computation overhead for individual users. Therefore, it is not well adopted in reality.

Block chain has decentralized and non-tamper features, so it is ideal for deploying electronic auctions on it. Recently, many researches have focused on combining block chain with auction. Kosba et al. present Hawk [2], a framework for creating Ethereum smart contract on the block chain. Anyone can write a Hawk program without having to implement any cryptography, its compiler can automatically generate privacy-preserving smart contract. In the Hawk program, the data and the flow of money will be blinded to the public. Hawk also utilizes zero cash technology to hide user identity. Hawk uses ZKP to prove the honesty of the manager. But studies have shown that it will take a long time to produce proof using ZKP and deploying ZKP in smart contract is complex. Blass and Kerschbaum present Strain [19], a protocol to implement sealed-bid auction on the block chain. Strain protects the bid privacy against fully malicious parties. Strain also designed a two-party comparison algorithm executed between any pair of bids for calculate the auction results in cipher text. But the protocol requires multiple interactions between each participant, and the communication and computation overhead are very large for individual users. Snchez [3] propose Raziel, a system that combines SMC and ZKP cryptographic primitive to guarantee the privacy, correctness and verifiability of smart contract. Furthermore, the author presents that a smart contract owner can prove its validity and correctness without revealing any information about the source message by using ZKP.

## 3. PRELIMINARIES

### 3.1. Bilinear Pairing

Throughout this paper, we will use this definition. Let $G_1$ be a cyclic additive group, whose orders is a prime $q$, and $G_2$ be a cyclic multiplicative group with the same order $q$. A bilinear pairing is a map $e: G_1 \times G_1 \to G_2$ with the following properties:

1. Bilinearity: $e(aP, bQ) = e(P, Q)^{ab}$ for all $P, Q \in G_1, a, b \in Z_q^*$.
2. Non-degeneracy: $e(P, Q) \neq 1$.
3. Computability: there exists an efficient algorithm to compute $e(P, Q)$.

### 3.2. Block chain and Smart Contract

A block chain can be referred to as a distributed database that chronologically stores a chain of data into sealed blocks [20] in a secure and immutable manner. Head-to-tail blocks guarantee that transactions are performed in an order, hence a transaction cannot be altered without changing its block and all the subsequent blocks. The content of the blocks can be written by the peers of the block chain through the consensus mechanism.

Block chain has four main properties [2]: 1) Reliable delivery of message. Because of the data written into the block cannot be modified. It is ideal regarded block chain as a ledger to ensure the persistence of message [21]. 2) Correct computation. The block chain can be seen as a state machine driven by transactions [22]. The miners continue to receive and validate new blocks, then package them on the chain, and the results of the calculations will be made public to all peers. 3) Transparency. All internal states and computations via the block chain will be visible to the whole block chain peers. 4) Pseudonym. A message or a transaction sends by one user in the block chain is referred to a pseudonym. The block chain address is usually generated by the user's public key.

In the Ethereum block chain [23], it provides the highest support for Turing's complete functionality by smart contract. They support the construction and execution of code that allow for the operation of a function on the block chain, which greatly enriches the flexibility of the block chain. Conceptually, a smart contract can be regarded as a special "TTP" [24], but this party is only for correctness and availability but not for privacy, because smart contracts deployed on block chain are also transparent.

## 4. TIME RELEASE PUBLIC KEY ENCRYPTION

The goal of time release public key encryption is to send an encrypted message to the future and wait until a specified time in the future to open. Let us assume that a sender wants to send a message to a receiver such that the receiver cannot be able to open it until a certain time. The encryption algorithm introduces a time server (referred to TS). The sender can encrypt the message using the public key of the recipient and TS without communicating with TS. Only after the release time has passed, the recipient can decrypt it by using his private key and the signature information (the information is related to the current time) from the TS. In addition, only the intended recipient holding the corresponding private key can recover the secret at some time (enforced by the trusted TS). So the time release public key encryption scheme is secure and private.

We describe a simple construction of time release public key encryption that is derived from the technology in [25]. The construction is based on bilinear mapping, and the security is based on the hardness of the Bilinear Diffie-Hellman Problem.

### 4.1. Time Release Encryption (TRE)

Suppose $G_1$ is additive cyclic groups, whose order is a prime $q$ and $G_2$ is multiplicative cyclic groups, whose order is also prime $q$. Let $e: G_1 \times G_1 \to G_2$ is a bilinear map. $G$ is a generator of $G_1$. Given the two cryptographic hash functions: $H_1: \{0,1\}^* \to G_1^*$; $H_2: G_2^* \to \{0,1\}^n$.

The TRE scheme contains five algorithms: (TS GEN, User GEN, TS broadcast, ENC, DEC), and it runs as follows.

**TS GEN:** The TS takes as input a secure parameters $k$ and outputs system parameters $params = \{k, q, G_1, G_2, e, G, H_1, H_2\}$ and key pair $(P_{TS}, S_{TS})$ of TS. The TS randomly selects $s$ as the private key $S_{TS}$, where $s \in Z_q^*$. Then TS computes $sG$ as the public key $P_{TS}$, $P_{TS} = (G, sG)$. Only $params$ and $P_{TS}$ are made public.

**User GEN:** Each user picks a secret key $a \in Z_q^*$ and computes the corresponding public key $(aG, asG)$.

**TS broadcast:** It runs by TS. TS inputs a time instant $T \in \{0,1\}^*$ and outputs a time-bound key update of the form $sH_1(T)$. TS automatically outputs the corresponding time-bound key for all current time instances T, the validity of which can be publicly verified by each user: checking the equation $e(sG, H_1(T)) = e(G, sH_1(T))$ is true, where $(G, sG) = P_{TS}$.

**ENC:** This algorithm is executed by sender. Given a message M, a receiver public key $(aG, asG)$, a $P_{TS}$, and a release time $T \in \{0,1\}^*$,

1) First, we need to verify whether the receiver really needs the server's time-bound key update message to decrypt the message M. So it checks $e(aG, sG) = e(G, asG)$; If the equation is true, the encryption algorithm continues.

2) Select a random number $r \in Z_q^*$, then compute $rG$ and $rasG$

3) Compute $K = e(rasG, H_1(T)) = e(G, H_1(T))^{ras}$

4) Output the cipher text $C = <U, V> = <rG, M \oplus H_2(K)>$.

**DEC:** This algorithm is executed by receiver. It inputs a cipher text $C$, a receiver private key $a$, and a time-bound key update $sH_1(T)$ from TS, that output is message M.

1) Compute $K' = e(U, sH_1(T))^a = e(G, H_1(T))^{ras} = K$.

2) Compute $V \oplus H_2(K')$ to recover message M.

### 4.1.1. A Sketch of Security Analysis

The security proof is the same to [25], we will simply describe it here.

1. The server private key is safe, for the Discrete Log (DL) problem is difficult (given $G, sG$, it is difficult to find $s$).

2. The user private key is safe, for this problem is at least as difficult as the DL problem (given $G, sG, aG, asG$, it is difficult to find $a$).

3. The server private key is safe, for that: to find $s$ from $\{G, sG, sH_1(T_1), sH_1(T_2),...\}$ to rewrite any $sH_1(T_i)$ is at least as difficult as the DL problem.

4. The decryption is difficult without having receiver and TS private key. If a receiver wants to decrypt a message before its release time, the easiest way is to solve the Bilinear Diffie-Hellman Problem (because the difficulty of the original problem is equal to the Bilinear Diffie-Hellman Problem in [25]). If the Bilinear Diffie-Hellman Problem is difficult, the receiver cannot decrypt any cipher text unless the release time arrival or he colludes with the TS.

## 5. BLIND SIGNATURE BASED ON ELLIPTIC CURVE

Blind signature [26] is a cryptographic protocol involving both the user and the signer. The user sends the blinded information to the signer, who signs the information but cannot obtain the specific content of the signed information. After the user receives the signed information and removes the blind factor, he can get the signature of the original message by the signer. Even if the signer sees this real signature, he cannot be sure if it came from his signature. Blind signature algorithm can effectively protect the specific content of signed messages or documents, so it plays a key role in the application of anonymity in electronic auction. Our protocol makes an extensive use of Blind signature scheme [27] which base on elliptic curve, and it has strong anonymity.

Common parameters are:

$E(F_q)$ : an elliptic curve defined on a finite field;

$G \in E(F_q)$ : a base point in elliptic curve;

$q'$ : a prime number;

$d \in_R Z_n^*$ : a signature private key; $Q = dG$ is a public key for verify the signature.

SHA-1: $\{0,1\}^* \to \{0,1\}^{160}$ is a cryptographic hash function.

Among the above parameters, $d$ is private and other parameters are public. Next we will describe the algorithm, where the notation $(\cdot \| \cdot)$ indicates to connect two bit strings, and $R_X(A)$ represents the coordinates of point $A$.

**SIG:**

1) The signer generates private key $k \in_R Z_n^*$, then calculates the corresponding public key $Y = kG$ and announce it to user.

2) The user picks three blind factors $\alpha, \gamma, \delta \in_R Z_n^*$, then calculates:
$A = \alpha Y + \gamma G + \delta Q = (x, y)$; $t = x \bmod n$; $c = SHA-1(m \| t)$; $c' = \alpha^{-1}(c - \delta)$, Where $m$ is the original message and $c$ is the blinded message. User sends $c'$ to the signer.

3) Signer calculates: $s' = k - c'd \bmod n$, where $s'$ is the result after the signer signs $c'$. The signer then sends $s'$ to the user.

4) User calculates: $s = \alpha s' + \gamma \bmod n$. $(c, s)$ is a blind signature for $m$.

**VER:**

The verifier checks if $c = SHA-1(m \| R_x(cQ + sG) \bmod n)$, if this equation is true, the signature is valid. Otherwise the verifier rejects the signature.

**A Sketch of Security Analysis:**

The specific security proof process can be referred to [27]. The validity of the signature is based on the security guarantee of the Schnorr Blind Signature Scheme, and the blindness is based on the DL Problem of the elliptic curve.

## 6. BLOCK CHAIN AUCTION PROTOCOL

### 6.1. System Model

In this section, we illustrate the specific process of the auction detail. Our system comprises four types of entities, as shown in Figure 1. The CA is responsible for issuing certificate to each user who is permitted to participate, and issuing public and private key pairs for two smart contracts. The auctioneer is responsible for announcing an auction task, publishing the list of users who are allowed to join in auction, the public parameters to be used in the calculation, the registration time, the bidding time instance and the finish time. During the auction process, the auctioneer also needs to sign the bid message for users, and finally decrypt all the bidding cipher text. The bidder bids in cipher text. Contract-1 and Contract-2 are deployed on the Consortium Block chain.

The serial number in the Figure 1 indicates the flow of the protocol:

1. Bidder generates the key pair $(x_i, y_i)$; 2. Bidder $B_i$ applies for registration from CA; 3. CA checks the bidder real identity, then issues $cert_i$ for $B_i$; 4. CA sends $(X_1, Y_1), (d, Q), (X_2, Y_2)$ via secure channel to the auctioneer; 5. Bidder blinds the bid $b_i$ to $c_i$; 6. Bidder applies for signature; 7. After the contract-1 verifies the $cert_i$ for $B_i$, it sends $c'_i$ to auctioneer; 8. The auctioneer sends the signature $s'_i$; 9. Bidder downloads the $s'_i$ from contract-2; 10. Bidder removes the blind factors; 11. Bidder sends the encrypted submission (bidding message); 12. The

contract-2 collects submission and sends $Esub_i$ to the auctioneer; 13. Auctioneer verifies the signature, decrypt the $Esub_i$, at last, submit the result on the chain.

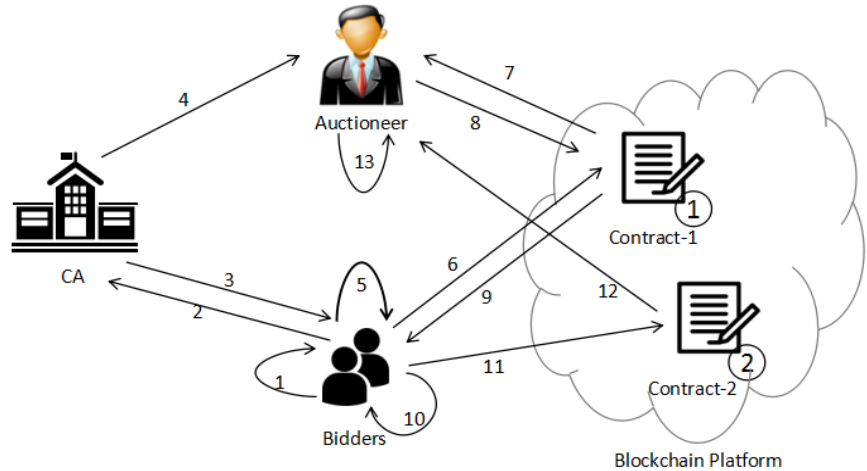

Figure 1. A process model of our auction protocol

We adopt consortium block chain in our scheme. Auctioneer communicates with bidders and smart contracts through an Ethereum block chain network where bidders send signature requests or encrypted bids to contracts, etc. After the contract receives parameters, the corresponding function will automatically execute and the execution result will be written into the block chain. Storing a private key in a smart contract is not secure, and the smart contract requires an external trigger can run. In order to prevent the auctioneer decrypt bids and leaked them in advance, as we mentioned before. We introduced a time release encryption algorithm when encrypting the bids, the auctioneer can decrypt the bids cipher text until the finish time arrival. This prevents the auctioneer from colluding with the bidder during the process of auction.

### 6.2. Definitions

List1: signature record table. The contract-1 records the signature information for each anonymous bidder to prevent bidder from bidding multiple times in one auction.

List2: bidding record table. The contract-2 stores the bidding information.

List3: it stores results of an auction task.

Contract-1: It stores the List1. The message sent to the contract-1 address in the form of a transaction through block chain network.

Contract-2: It stores List2 and List3. The message sent to the contract-2 address also in the form of a transaction.

### 6.3. The Construction of Protocol

#### 6.3.1. Parameters Setup and Auction Publish

The auctioneer initially setup the parameters to be used in an auction task, and broadcasts the auction task in the block chain network. TS also runs the TS GEN algorithm in TRE protocol as mentioned before to initialize parameters.

1) Setup the auction task and deposit the budget. The auctioneer deploys contract-1 and contract-2. The deposit is a sum of money that the auctioneer needs to send into the contract-2 account. Assume the auctioneer has aborted the protocol or has been caught cheating, then the money will be distributed to the bidders as a punishment.

2) $T_1, T_2, T_3, T_4$ define the time intervals for the following four phases: register at CA, sign for bids value, bidder submits bid, and publish the auction result, respectively.

3) CA picks elliptic curve $E(F_q)$ and the base point $G \in E(F_q)$.

4) TS runs the TS GEN algorithm to initialize and announce system parameters $params$, $P_{TS}$, where $P_{TS}$ is the public key of TS. The auctioneer runs the User GEN algorithm in TRE protocol to generate a key pair $(aG, asG)$ (which will be used for bidders to encrypt bids) for this task only.

5) The auctioneer, CA and TS announce the above public parameters and informs the bidders to start registration. The registration phase should be completed within $T_1$.

#### 6.3.2. Register at Certification Authority

The bidder registers at CA to get a certificate bound to his/her unique identity ID. At the same time, CA generates public and secret key pairs for contract-1 and contract-2.

1) The bidder $B_i$ generates a random number $x_i$ as a secret key and calculates the corresponding public key $y_i = x_i G$.

2) $B_i$ sends $\{y_i, ID_i\}$ to the CA.

3) The CA checks the identity of the bidder $B_i$ and checks whether he is eligible to participate in this auction. After the review is passed, CA will issue a certificate $cert_i$ to $B_i$.

4) CA generates signature key pairs for contract-1, $(X_1, Y_1)$ and $(d, Q)$, then sends $X_1, d$ to the auctioneer over the secure channel. The secure channel of this paper is implemented by TLS, which ensures the confidential and integrated for information, and also it can prevent eavesdropping.

5) The contract-1 generates the signature record table List1. The list consists of the bidder's certificate and the bid flag. The flag is used to indicate whether the user has applied for a signature before, 0 means no application has been made, and 1 means already applied. List1 is shown in Table 1.

6) CA generates the key pair of contract-2 $(X_2, Y_2)$, and sends $X_2$ to the auctioneer through the secure channel.

Table 1. Heading and text fonts

|   | **Bidder's Cert** | **Bid Flag** |
|---|---|---|
| 1 | $Cert_1$ | 0 |
| 2 | $Cert_2$ | 1 |
| n | $Cert_n$ | 0 or 1 |

### 6.3.3. Sign for Bids Value

The bidder blinds the bid that needs to be signed, then he sends it to the contract-1 address. After verify it by List1, the auctioneer will collect the application from contract-1 and sign it, then send the signature back to the contract. The user can get his signature directly from contract-1.

1) Bidder $B_i$ selects the blinding factors $\alpha_i, \gamma_i, \delta_i \in_R Z_n^*$, then calculates:
$A = \alpha_i Y_1 + \gamma_i G + \delta_i Q = (x_i, y_i)$; $t_i = x_i \bmod n$; $c_i = SHA-1(b_i \| t_i)$; $c_i' = \alpha_i^{-1}(c_i - \delta_i)$, Where $b_i$ is the plaintext of bid value, $c_i$ is the blinded bid value.

2) The bidder sends $\{cert_i \| c_i'\}$ to the contract-1 via the secure channel.

3) After contract-1 receives the blinded bid, it checks whether the bidder's certificate is legal. If the validation fails, it will refuse to sign for the blinded bid.

4) The contract-1 will check if the flag corresponding to $cert_i$ in List1 is equal to 0, that is, check if the user has applied for a signature before. If the flag is 1, then it refuse to sign.

5) The auctioneer keeps the secret key $d$ of contract-1, so he will download the $\{cert_i \| c_i'\}$ and execute signing, that is, calculates: $s_i' = k - c_i'd \bmod n$. Then it marks the position of $cert_i$ in List1 as 1.

6) After the auctioneer sends the signature $s_i'$ to contract-1, bidder $B_i$ can get the signature $s_i'$.

### 6.3.4. Bidder Submits Encrypted Bid Value

The bidder recovers the signature of the original bid value by removing the blind factor, then he encrypts the bid value and sends it to the contract-2 anonymously.

1) $B_i$ removes blind factors in $s_i'$, that is calculates: $s_i = \alpha_i s_i' + \gamma_i \bmod n$. It can get the signature $s_i$ for the original bid value.

2) $B_i$ packs the submission $\{(c_i, s_i) \| b_i\}$, then he runs the ENC algorithm in TRE we mentioned before to encrypt it, that is calculated: $Esub_i = ENC\{(c_i, s_i) \| b_i\}_{Y_2, P_{TS}, T}$.

3) $B_i$ sends $Esub_i$ to the contract-2 by his one-time address.

### 6.3.5. Statistics and Publish the Results

TS periodically runs the BCST algorithm on all time instances, in order to calculate time-bound key $sH_1(T)$ and broadcasts it. The auctioneer will get the correct $sH_1(T)$ from the TS when the specified decryption time is reached. The auctioneer collects the cipher text of bids can decrypt them, and verifies the validity of the signature. The contract-2 also checks if the user has been bid before, for the List2 has stored the bidding record. Because the signature scheme we used has a strong anonymity feature, we can guarantee that user's certificate can be used only once in the same auction process. Then the auctioneer computes the auction result. The user with the lowest bid is winner, he will receive the auctioneer's deposit. Finally, the results will be written into the ledgers and published.

1) After all $Esub_i$ are sent to the contract-2 address, the auctioneer collects them and uses the secret key $X_2$ and a time-bound key update $sH_1(T)$ to decrypt them. For each $Esub_i$, it will run the DEC algorithm in TRE to calculate $\{(c_i, s_i) \| b_i\} = DEC\{Esub_i\}_{X_2, sH_1(T)}$ to get the real bidding message. Subsequently, it verifies that the validity of the signature $(c_i, s_i)$. That is to calculate: $c_i = SHA-1(b_i \| R_x(c_iQ + s_iG) \mod n)$. If they are equal, the signature is valid; otherwise, the $Esub_i$ will be discarded.

2) Pick out $c_i$ and check if there exits $c_i$ in the bidding record table List2. If it exists, discard it; otherwise, write $c_i$ into List2.

3) After all bids have been decrypted, the auctioneer collects $b_i$, then compares these bids to gets the auction results.

4) The auctioneer announces all $\{(c_i, s_i) \| b_i\}$ to contract-2 and adds the result into List3. At last, the $\{(c_i, s_i) \| b_i\}$ and List3 will be written into the ledger after the consensus algorithm and node verification.

In this process, if the auction activity cannot proceed normally due to the malicious auctioneer aborts the protocol, the contract will automatically allocate the auctioneer's deposit to the participating users. Due to the blindness of blind signatures, even if the auctioneer sees this real signature, he cannot be sure who it comes from. So he cannot link one user's bidding behavior to his true identity.

### 6.4. Analysis of the Protocol

**Financial fairness and correctness.** It is clear to see the auctioneer will obtain the result after certain time, the winner would receive the amount of payments. Condition on that they all follow the protocol, the block chain can be modeled as an ideal public ledger, the time release public key encryption is correct and confidential and the blind signature based on elliptic curve with strong anonymous and correctness.

**Non-repeatable bid.** The malicious bidders is straightforward, the way they can cheat are: (i) submitting more than one bids for one $(c_i, s_i)$ in section 6.3.4; (ii) steal other bid information in advance; (iii) sending the contract-2 a fake instruction without having a legal identity in bidding submitting phase; (iv) altering the policy specified in the contract. The first threat is simply handle by the blind signature based on elliptic curve with strong anonymous, it can ensure one identity bind one bid in an auction. The second treat can be prevent due to the correctness and security of time release public key encryption. The third threat can be simple handled by the contract-1 and contract-2 store the List1 and List2, and the digital signature is security. The last threat is trivial, for the block chain security ensures the data on the chain is immutable.

**Bid privacy.** If the malicious auctioneer deny the payments or other policy announced in auction publish phase (section 6.3.1) or decrypt and leak the bids in advance. The first issue is prevented because the smart contract is publicly on the chain. The second threat is prohibited by the fairness of TRE protocol, the auctioneer can decrypt these bids until certain time arrival.

**Identity anonymity.** Every auctioneer/bidder will generated one-task-only block chain address with each auction activity, and the blind signature scheme hidden the user identity. Only the CA and auctioneer know who participate in this auction, but no one can bind a certain bid to a unique identity, and the bidding behavior or the same user in multiple auctions cannot be linked.

**Public verifiable correctness.** The auction result with each user's bid message will be open without disclose their identity. For a bidder, he can check his bid has been recorded on the chain. For the auction results are public, everyone can verify the correctness. Subsequently, the winner can find the auctioneer privately for follow-up trade.

**Theorem 1.** The bids confidentiality of our protocol holds, condition on that the underling time release public key encryption is semantically secure and the TS is honest. The anonymity of our protocol for bidders will be satisfied, if the underlying blind signature based on elliptic curve satisfies the strong anonymity and blindness. If the Consortium Block chain infrastructure we rely on can be regarded as an ideal public ledger, the underlying encryption and blind signature protocol are secure, our protocol satisfies: security against a malicious auctioneer and against malicious bidders.

## 7. DISCUSSION

We compare our scheme to some existing auction schemes deployed on the block chain. For the privacy of scheme, Strain [19] propose to use pseudonym in hiding user identity, but when the first money is transferred to a new address, his identity will be exposed. Hawk [2] utilizes Zero cash technology to hide identity, which particularly used in crypto currencies. It uses ZKP to prove the honesty of the manager, so it must deploy the ZKP function on the smart contract. Raziel [3] combines SMC and ZKP cryptographic primitives to guarantee the identity privacy, correctness and verifiability of smart contract. The main challenge of deploying smart contracts is that they can only support very light operations for computing. It is more powerful, but the deployment of the experiment is more complicated. According to the experimental results in the Hawk, it takes at least hundreds of seconds to generate proof using ZKP in auction scenarios.

International Journal of Advanced Information Technology (IJAIT) Vol. 9, No.1, February 2019

We use the blind signature based on elliptic curve, one-time block chain address and the strict identity access mechanism accompanied by Consortium Block chain to guarantee the privacy of the participants and smart contract. Moreover, we are the first to introduce the TRE technology to prevent auctioneer from leaking bidding information in advance and collusion with bidders. We use smart contracts to store and collect data, without having to perform very complex operations, so it takes very little time for the contract to run on the chain.

As far as the computation/communication overhead on-chain. For the contract-1, it only stores applications for the signatures from user and maintains the List1. For the contract-2, it only collects the cipher text of bidding information (submissions) and maintains the List2. The main computational cost is borne by the auctioneer, and he should sign the bidding information and decrypt the cipher text of bids. But this calculation can also be outsourced to a secure computing center, or an external hardware to deploy cryptographic functions if necessary. For the participating supplier, their local computing and communication overhead is also very small. Therefore, the on-chain performance of the system can be clearly practical regarding time, and the entire protocol will be efficient.

## 7.1. Performance Evaluation

We implement our auction protocol atop Ethereum, and conduct experiments of auction tasks in an Ethereum test net to evaluate the applicability. We simulated several types of nodes on the chain to run the entire protocol process. Experiment results have shown that our protocol is feasible, and it spends little time on the chain, as well as meets the requirements mentioned earlier.

On the other hand, we have measure the main computation costs for auctioneer and bidder in our protocol with Python, the experiments are conducted on Ubuntu 16.04.3 LTS with Intel(R) Core(TM) i7-6500 CPU 2.50GHz and 4GB RAM. To achieve better accuracy, we have performed the blind signature test 500 times and the TRE test 500 times, and we choose the average value of all results. All nodes in the private p2p test network were connected and propagated transactions and messages to each other. On the testnet we generated our own private block chain creating unique Merkle hash root and new block. Because of the smart contracts can only support very light on-chain operations for computing, the heavy computation of signature and encryption are done off-chain. In our implementation, we increased the number of nodes for testing. We set other parameters by default, for instance, the length of data is 100 bits, the length of random number is 400 bits, the length of modulus is 512 bits and the length of exponent is 80 bits. Then we tested the performance of each algorithm with different number of users. The performance of each algorithm was shown in Figure 2,3, the algorithm included "ENC" and "DEC" in TRE protocol, the "blind", "sign" and "verify" in blind signature protocol.

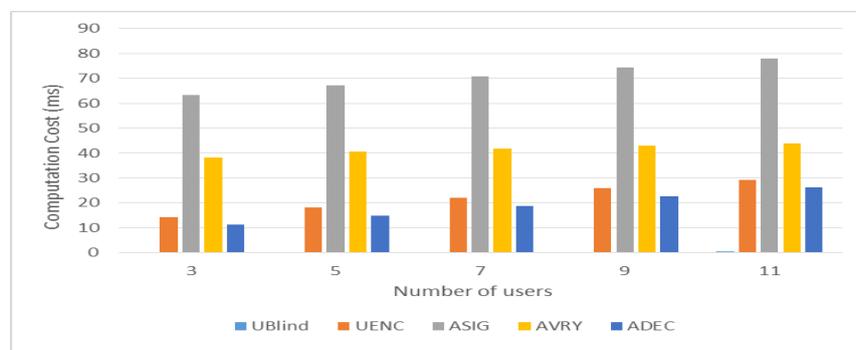

Figure 2. Computation time of each algorithm in our protocol.

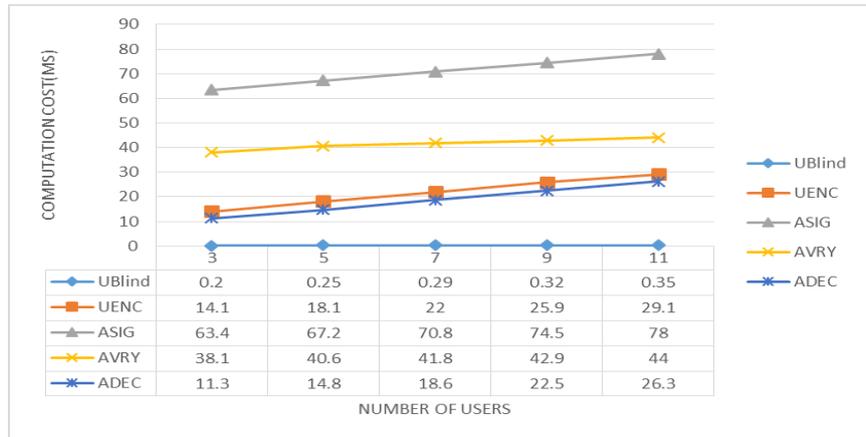

Figure 3. Computation time with majority bidders.

Notes: UBlind denotes the blind signature algorithm used in our paper runs by a bidder; UENC denotes the TRE encryption algorithm runs by a bidder; ASIG denotes the signature algorithm runs by an auctioneer, AVRY denotes the verification algorithm runs by an auctioneer and ADEC denotes the TRE decryption algorithm runs by an auctioneer.

In general, the above tests show that we have achieved the lightweight of Blind signature and TRE in our scheme, nearly all of the computation overhead has been transferred to user off-chain. It is clear that the time to store the date on the chain can be efficiently executed, therefore, our auction protocol on the block chain can be clearly efficient and feasible.

### 7.2. Future Work

As the current smart contract technology is at early stage and can only allow deploy very easy operation or allow tiny storage. Can we further explore the possibility of allowing smart contract to automate operations such as decryption and verification of signature? On the other hand, can we use new storage technologies, or combine off-chain storage to assist more large scale auction activities. For the anonymity, we can investigate other approaches applicable on consortium block chain where we can protect the privacy of the auctioneer.

### 8. CONCLUSIONS

In this paper, we presented an anonymous sealed-bid auction protocol on the consortium block chain. We adopt a strict digital certificate-based identity mechanism of the consortium block chain to permit legitimate participants. The auction protocol maintains the privacy of bids so the bidders do not know any information about the other bid before certain time, and we achieve it through a time release encryption scheme. Additionally, we use the strong blind signature based on elliptic curve technology to protect the privacy of the bidder identity. Moreover, the auction smart contract have no complex operations, the computation overhead on the chain is very tiny.